# Refractive index measurements in absorbing media with white light spectral interferometry

YAGO AROSA, ELENA LÓPEZ LAGO, AND RAÚL DE LA FUENTE*

*Grupo de Nanomateriais, Fotónica e Materia Branda, Departamentos de Física Aplicada e de Física de Partículas, Universidade de Santiago de Compostela, E-15782, Santiago de Compostela, Spain*
*raul.delafuente@usc.es*

**Abstract:** White light spectral interferometry is applied to measure the refractive index in absorbing liquids in the spectral range of 400–1000 nm. We analyze the influence of absorption on the visibility of interferometric fringes and, accordingly, on the measurement of the refractive index. Further, we show that the refractive index in the absorption band can be retrieved by a two-step process. The procedure requires the use of two samples of different thickness, the thicker one to retrieve the refractive index in the transparent region and the thinnest to obtain the data in the absorption region. First, the refractive index values are retrieved with good accuracy in the transparent region of the material for 1-mm-thick samples. Second, these refractive index values serve also to precisely calculate the thickness of a thinner sample (~150 μm) since the accuracy of the methods depends strongly on the thickness of the sample. Finally, the refractive index is recovered for the entire spectral range.

## References and links

## 1. Introduction

Basic characterization of a homogeneous optical material includes the determination of its linear optical constants: the refractive index and linear absorption coefficient. Although these quantities are closely related, they are frequently measured independently. Abbe refractometry [1], ellipsometry [2], or the method of minimum deviation [3] are optical techniques often used to measure the refractive index. However, these techniques provide the values of the refractive index only at a given wavelength or discrete set of wavelengths and require interpolation or empirical formulas to obtain the refractive index in a broad spectral band. On the other hand, absorption spectrometry [4] makes use of a broadband source to measure absorbance or the absorption coefficient in a wide range of wavelengths (Vis, near-UV and/or near-IR). The measurement of the refractive index in such a wide spectral range is a difficult task, especially if high-precision measurements are required (with uncertainties of $\sim 10^{-4}$). However, the combination of white light interferometry and spectroscopy has been proved to provide precise refractive index values in the Vis range and beyond [5,6]. This technique, named white light spectral interferometry (WLSI) or, alternatively, spectrally resolved white light interferometry, has been applied successfully to measure displacement and thicknesses [7,8], dispersion in transparent samples with an accuracy of $\sim 10^{-4}$ [9], group delay dispersion on the infrared range between 2 and 20 μm [10], chromatic dispersion of polarization modes of optical fibers [11], optical phase measurement in metamaterials [12]. This technique was used by Rocha et al [13] to extract the thermo-optical coefficient on a broad continuous visible range with an accuracy of $\sim 10^{-3}$. Yet, absorption makes it difficult to use this technique, not only due to the decreased detected signals but also due to the reduction in the visibility of fringes in the spectral interferogram. This is a main problem when characterization materials with high absorbance in the working window, like semiconductors for solar systems [14]; color filters for displays [15], medical applications to analyze blood and tisues [15,16], optical systems with high absorption, like plasmonic interactions [17] or turbid colloids` [18]. One way to overcome this limitation is to use thinner samples to lower the absorbance. However, in transmission measurements, knowledge of the sample thickness is required to calculate the refractive index, and, unfortunately, the refractive index uncertainty increases with decreasing thickness.

This paper presents broadband refractive index measurements by WLSI in absorbing media; we analyze the influence of absorption in the measurement and discuss its limits; we further propose a two-step process as a means to overcome these limits. In the first step, the refractive index is measured in the transparency window of the material using a thick sample. Next, the measured refractive indices are used to calculate the thickness of a thinner sample. Once the sample thickness is obtained with high accuracy, the refractive index in the absorption region is determined with a similar accuracy to that measured for a transparent material.

## 2. Background

WLSI uses a Michelson interferometer illuminated by a broadband source to obtain at its output the superposition of light waves travelling along different paths in the interferometer arms. Since the optical path length is inversely proportional to the wavelength of light, the total irradiance at the output of the interferometer is incoherent superposition of the individual interference patterns for each wavelength and does not carry any information about the optical path length difference. However, we can separate different spectral components with an optical spectrometer to obtain the output irradiance as a function of wavelength:

$$I(\lambda) = I_1(\lambda) + I_2(\lambda) + 2\sqrt{I_1 I_2}\, v_s(\lambda) \cos\varphi(\lambda), \tag{1}$$

where $I_1$ and $I_2$ are the irradiance values in the two arms, $v_s$ is the spectrometer visibility function defined by its limited bandwidth and resolution, and $\varphi$ is the phase related to the optical path length difference. When a sample of width $d$ and absorption coefficient $\alpha(\lambda)$ is placed in one arm of the interferometer (Arm 1), we can express the irradiances in the interferometer arms in terms of the sum, $I_0$, and the difference, $\Delta I$, of the irradiances without absorption as

$$\begin{aligned} I_1 &= \frac{1}{2}(I_0 - \Delta I)e^{-2\alpha d} \\ I_2 &= \frac{1}{2}(I_0 + \Delta I), \end{aligned} \tag{2}$$

where we have taken into account that the optical beam in Arm 1 propagates through the absorbing sample twice. On the other hand, the phase difference is written as

$$\varphi(\lambda) = \frac{4\pi}{\lambda}\big[d(n-1) - L\big], \tag{3}$$

where $n(\lambda)$ is the wavelength-dependent refractive index, and $L$ is the difference in length between the interferometer arms.

Using Eq. (2), we rewrite the total irradiance as

$$I(\lambda) = I_B(\lambda)\big[1 + v(\lambda)\cos\varphi(\lambda)\big], \tag{4}$$

where $I_B(\lambda)$ is the background irradiance:

$$\begin{aligned} I_B(\lambda) &= I_1(\lambda) + I_2(\lambda) \\ &= I_0(\lambda) e^{-\alpha d}\big[\cosh(\alpha d) + \Delta \sinh(\alpha d)\big], \end{aligned} \tag{5}$$

and $v(\lambda)$ is the total visibility function, given by

$$v(\lambda) = \frac{\sqrt{1-\Delta^2}\, v_s(\lambda)}{\cosh(\alpha d) + \Delta \sinh(\alpha d)}, \tag{6}$$

with $\Delta = \Delta I / I_0$. Thus, the effect of absorption is to decrease the background irradiance and, most importantly, to reduce the visibility function. When the visibility function decreases, it is more difficult to accurately extract the phase from the measured irradiance given by Eq. (4) and, hence, to determine the refractive index as a function of wavelength.

## 3. Experimental

In order to analyze this effect, we prepared solutions of a known lens colorant RED-06847 (CR39) in deionized water with different concentrations (see Table 1). Figure 1 shows the absorbance of each sample in a 2-mm quartz cell, measured with a commercial monochromator (Cary 100 Bio from Varian Inc.). The absorbance peaks at 535 nm and extends over the visible and NIR regions decreasing for longer wavelengths.

Table 1. Sample Concentrations

| Sample | Concentration [g/L] | Sample | Concentration [g/L] |
|---|---|---|---|
| 1 | 0.08 | 8 | 10.64 |
| 2 | 0.15 | 9 | 12.16 |
| 3 | 3.04 | 10 | 13.68 |
| 4 | 4.56 | 11 | 15.20 |
| 5 | 6.80 | 12 | 15.84 |
| 6 | 7.60 | 13 | 17.82 |
| 7 | 9.12 | 14 | 19.80 |

Figure 2 compares the interferogram obtained for sample 11 with the one obtained for the pure deionized water sample. The envelope changes drastically while the oscillations remain practically the same (the variations are due to the change in $L$ occurring when changing the sample). The visibility is greatly reduced for wavelengths shorter than that of the stationary phase point ($\lambda = 620$ nm), making the determination of the refractive index in this spectral region extremely difficult.

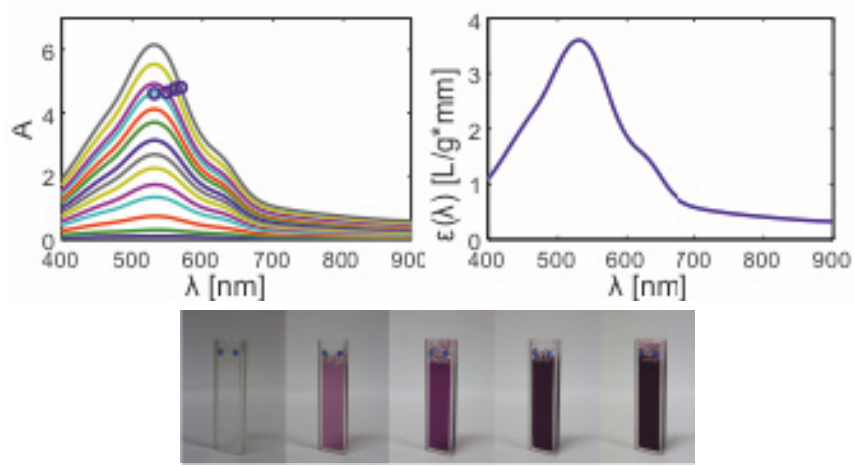

Fig. 1. (a) Absorbance of solutions of CR39 dye in deionized water, $A = -\log T = 0.433\alpha d$ with $d = 2$ mm; the circles correspond to the cut-off wavelengths (see Table 2 and the related text). (b) Absorptivity, $\varepsilon = \alpha/c$. (c) Photographs of samples 1, 3, 6, 9, and 12 inside a 1-mm-thick cell.

It is possible to extract the background irradiance and the visibility function from the interferograms by interpolating between the maxima and minima to extract the upper and lower envelopes. Special care must be taken near the stationary phase point because of the lack of fast oscillations that are required for extracting the visibility with sufficient accuracy. In Fig. 3, the background irradiance and visibility are shown as functions of absorbance at its peak ($\lambda = 535$ nm) together with the corresponding theoretical values. The values of $\Delta$ and $v_s$ were obtained by fitting Eqs. (5) and (6) as functions of $e^{-\alpha d}$ for this wavelength. In the figure, it is seen that the visibility is only a few per cents for the absorbance of ~7 or more. This defines the spectral region where the phase and refractive index cannot be retrieved with sufficient accuracy.

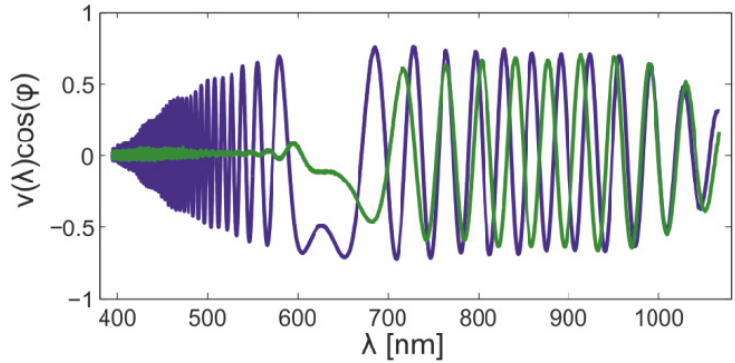

Fig. 2. Interferograms for pure deionized water (blue line) and for solution 11 (green line) after subtraction of the background irradiance.

To define the aforementioned region more precisely, we now consider the extraction of the phase in Eq. (4) and retrieval of the refractive index. First, let us recall that the arc cosine is a multivalued function, and so is the phase obtained from Eq. (4); hereafter, $\varphi_{exp}(\lambda)$ differs from the phase in Eq. (3) by an unknown multiple of $2\pi$, that is:

$$\varphi_{exp}(\lambda) = \varphi(\lambda) - 2k\pi = \frac{4\pi}{\lambda}\left[d(n-1) - L\right] - 2k\pi. \qquad (7)$$

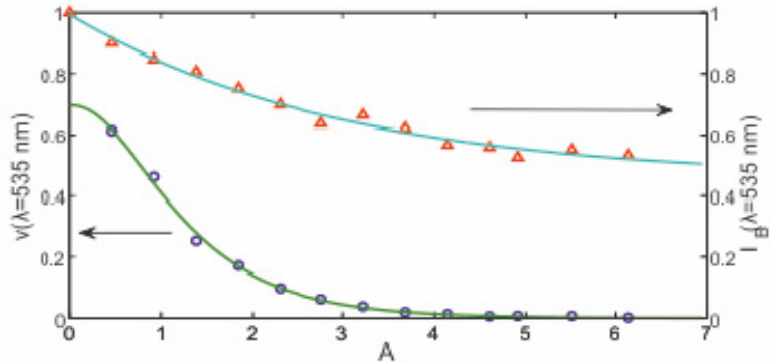

Fig. 3. Relation between visibility, normalized background irradiance, and absorbance. The circles and triangles indicate the values obtained from the interferograms, while the continuous curves represent theory.

In order to compare the different phases, we must first measure $L$ for each sample (using WLSI as well [8]) and subtract it in Eq. (7); then, the factor $k$ can be eliminated by arbitrarily selecting the phase to be in the interval $[-\pi, \pi]$ for a given reference wavelength. The correct extraction on these factors has been deeply studied in a previous work [12] where the challenges of the technique to achieve the desired uncertainty are described in detail; it has been demonstrated that it is a mandatory requirement that the value of $L$ be retrieve with enough accuracy (errors of $\sim 10^{-4}$ or less) to achieve an uncertainty on the fourth decimal figure of the refractive index. Since we are using liquid samples; it must be noted that to compensate the dispersion of the glass cells, it is necessary to place an identical empty cell in the other interferometer arm. It has been also shown that as thinner is the sample the accuracy on the retrieval decreases. In Fig. 4, we plot the different phases and their differences with the one obtained for the pure deionized water sample. The difference is less than 2 radians in the entire measured spectral range for the first ten solutions (they are widely spaced at shorter wavelengths); this corresponds to a difference in the refractive index of less than $10^{-4}$ at 500 nm. This means that the dye at low concentrations only adds to the absorption and leaves the refractive index unaltered (at least with this level of precision). Furthermore, the refractive index can be determined as in a transparent sample. However, for samples 11 to 14, the phase begins to differ significantly from the one corresponding to pure deionized water and reaches tenths of radians at the shortest wavelengths.

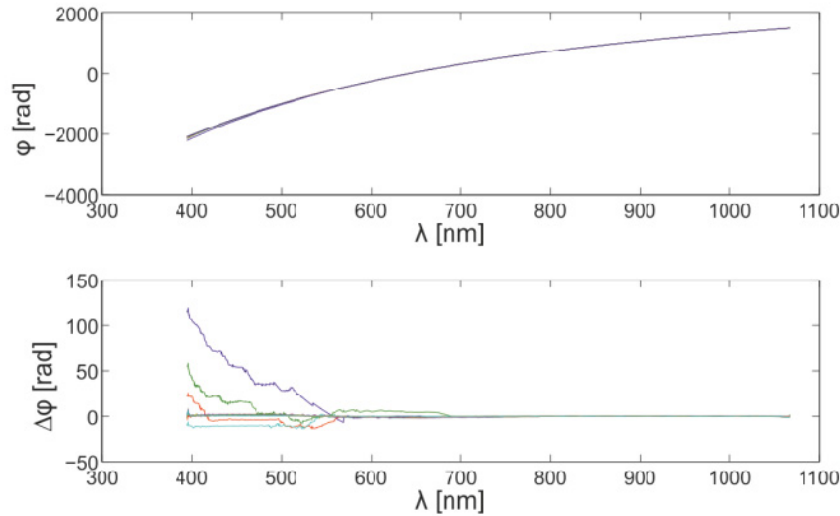

Fig. 4. a) Phases obtained for different solutions; b) difference of the obtained phases with that obtained for the deionized water sample.

As a consequence, for the last four samples, the refractive index is calculated with insufficient accuracy in the short-wavelength part of the spectrum. We can define a cut-off wavelength, $\lambda_c$, which limits the region where the refractive index is correctly determined. We can define this limit arbitrarily by choosing $\lambda_c$ as the wavelength where the phase difference induces a change on the refractive index as large as $10^{-4}$. Then, to obtain $\lambda_c$, we must solve numerically the following transcendental equation:

$$\frac{\lambda_c \Delta\varphi(\lambda_c)}{4\pi d} = 10^{-4}. \qquad (8)$$

In Table 2, we show the cut-off wavelength together with the sample absorbance and interferogram visibility at this wavelength for the last four samples listed in Table 1.

Table 2. Cut-off Wavelength, Corresponding Absorbance, and Visibility of the Interferogram

| Sample | $\lambda_c$ | $A(\lambda_c)$ | $v(\lambda_c)$ |
|---|---|---|---|
| 11 | 532 | 4.61 | 0.007 |
| 12 | 550 | Sample | 0.008 |
| 13 | 561 | 4.76 | 0.007 |
| 14 | 569 | 4.80 | 0.005 |

The cut-off wavelength belongs to the green region of the spectrum in all the cases and shifts to longer wavelengths as the peak absorption of the sample increases. Meanwhile, the absorbance increases in the same way and varies around a mean value of 4.7, with the visibility of ~0.7%. We find that WLSI works very well at very low values of visibility. It must be noted that, in addition to absorbance, the total visibility depends of the difference in spectral irradiance of the signals travelling in the two interferometer arms (given by $\Delta$ in Eq.

(6)) and the spectrometer visibility, $v_s$; thus, for a fixed value of the total visibility, the absorbance at the cut-off wavelength increases as long as $\Delta$ or $v_s$ decrease, and vice versa.

Although one obvious way to expand the spectral range where accurate results can be achieved is to reduce the sample thickness, $d$, and, accordingly, mitigate the absorbance, it must be noted that the refractive index uncertainty grows as $1/d$ ($\Delta n/n = \Delta d/d$). Thus, in order to improve the refractive index uncertainty we need minimize the uncertainty in $d$, We propose to perform the measurement in two steps. In the first step, WLSI is applied to retrieve the refractive index in the transparency window of the material. In the second step, the measurement is taken for a thinner sample. The thickness of this sample is calculated with sufficient accuracy by using as reference the retrieved refractive indices in the transparent window [9]; once $d$ is known, we recover the refractive index in the remaining spectral range. To understand how $d$ is calculated, we rewrite Eq. (7) as

$$\frac{\varphi_{\exp}(\lambda)}{2\pi} - \frac{2L}{\lambda} = 2d\frac{(n-1)}{\lambda} - k. \tag{9}$$

All the parameters in the left-hand side are known. In the right-hand side, only $d$ and $k$ are unknown. They can be determined by taking $y = \varphi_{\exp}(\lambda)/2\pi - 2L/\lambda$ and $x = 2(n-1)/\lambda$ and performing linear least square fitting.

Table 3. Values of *k* obtained from Linear Fitting, Rounded Values, and Final Fitted Thicknesses of the Thin Samples

| Sample | $k_1$ | $k_2$ | $d$ (μm) |
|---|---|---|---|
| 11 | 19.95 ± 0.14 | 20 | 154.30 ± 0.13 |
| 12 | 22.06 ± 0.23 | 22 | 131.59 ± 0.15 |
| 13 | 21.09 ± 0.16 | 21 | 147.70 ± 0.16 |
| 14 | 20.98 ± 0.15 | 21 | 164.75 ± 0.14 |

We used ~150-μm-thick samples, constructing cells with two quartz windows separated by microscope cover slips. These cells reduce the absorbance of the formerly used cells by ~1/7. The spectral range of 700–1000 nm, where all the thick samples are transparent, is used to perform the fit. The results are shown in Table 3. We note that the obtained values of $k$ are nearly but not exactly integer. Since this is a necessary condition, we round $k$ to the nearest integer and perform a second fit with $k$ fixed, to obtain the final value of $d$. We find that the value of $d$ varies from sample to sample. This is due to the variation of the cover slip thickness.

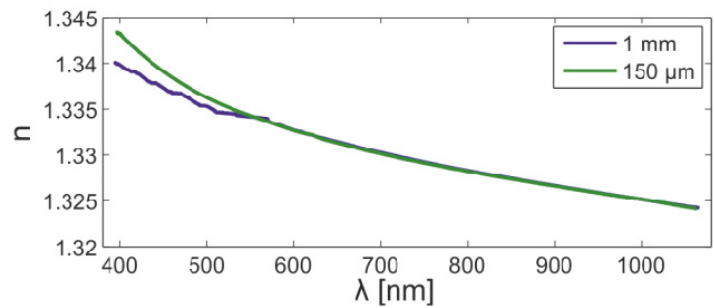

Fig. 5. Refractive indices for solution 14. The cells used are 1-mm thick (blue line) and 150-μm thick (green line).

Figure 5 shows the refractive index curves obtained for the two cells for sample 14. As noted, the results for the thickest cell are not accurate in the spectral range of 400–570 nm; on the other hand, we are able to obtain accurate refractive indices in the entire spectral range for the thinnest cell.

As another example, we consider the ionic liquid tris(1 butyl-3 methylimidazolium) tetra isothiocyanate cuprate, $[BMIM]_3[Cu(SCN)_4]$ [19]. Ionic liquids are defined as ionic compounds that remain liquid on the range of temperatures between 0 and 100 °C. Their properties can be tuned using different combinations of cation and anion with the possibility of doping them for further modulation of such characteristics. A proper optical characterization is mandatory when using dopants since they usually change the absorption profile of the liquid as well as the refractive index. The presence of copper in this ionic liquid results in additional absorption in the blue region of the spectrum, as shown in Fig. 6. We were only able to recover the refractive index curve above 550 nm with the 1-mm cell. From 550 to 1000 nm, the retrieved refractive index was well fitted with the Cauchy-type dispersion formula:

$$\begin{aligned}n^2 - 1 = &\ 1.5\,10^{-3}/\lambda^4 + 2.08\,10^{-2}/\lambda^2 \\ &+ 1.3901 - 7.6\,10^{-3}\lambda^2 + 2.3\,10^{-3}\lambda^4,\end{aligned} \qquad (10)$$

where the units of wavelength are microns. When using the 150-µm-thick cell, the refractive indices were successfully retrieved for wavelengths higher that 460 nm. Hence, we must use an even thinner cell to obtain the refractive index in the far blue part of the spectrum. However, as shown in Fig. 6(c), Eq. (10) fits the experimental refractive index with a precision of better than $10^{-4}$ up to 460 nm, although the fit was performed only from 550 to 1000 nm. Therefore, it is expected that it can be extrapolated to the spectral range of 400–460 nm.

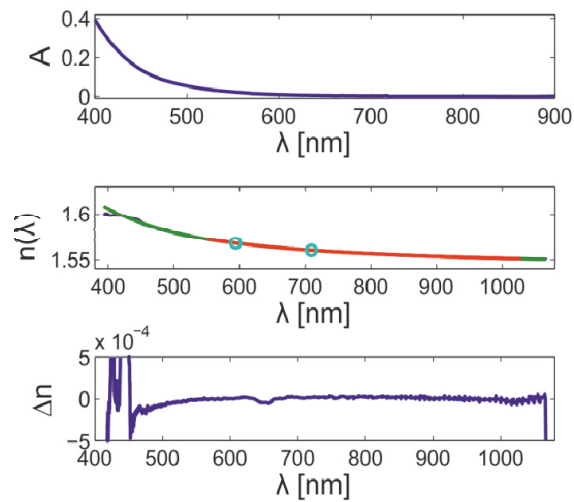

Fig. 6. (a) Absorbance of $[BMIM]_3[Cu(SCN)_4]$ for a 10-µm-thick cell, (b) refractive index obtained with a 1-mm-thick cell (red), refractive index measured using microscope cover slips (blue), and fitted refractive index (green); the spectral range used to extract $d$ is the one between the circles; (c) difference between the refractive index given by Eq. (10) and the experimental refractive index obtained with a 150-µm-thick cell.

## 4. Conclusions

In conclusion, WLSI, a well-known technique used to determine refractive indices in transparent materials, was proved to work properly with absorbing liquids, achieving similar levels of accuracy. Using dilute CR39 dye solutions in deionized water, we verified that the refractive index can be retrieved with sufficient accuracy while the visibility of fringes exceeds 1%, a surprisingly small value; this implies values of absorbance smaller than 4.6 and the absorption coefficient lower than 53 cm$^{-1}$ when using a 1-mm-thick cell and 353 cm$^{-1}$ when using a 150-μm-thick cell. The 1-mm-thick cell serves to retrieve the refractive index in the transparent region and also to determine with enough accuracy the width of the thinnest cell and to guarantee a good refractive index retrieval in the absorbing region. Furthermore, we demonstrated that the possibility of determining precisely the lengths or thicknesses of samples with WLSI in dispersive media extends the spectral range where the refractive index can be retrieved. Although we used only liquid samples to test the technique, the results presented here apply to solid absorbing media as well.

## Funding

Ministerio de Economía y Competitividad (MINECO) (MAT2014-57943-C3-2-P, MAT2017-89239-C2-1-P); Xunta de Galicia and ERDF (AGRU 2015/11, GRC ED431C 2016/001 and ED431D 2017/06).

## Acknowledgment

We thank Julio Seijas from the Synthetic Weird Technology Research Group of the USC who prepared and provided the ionic liquid [BMIM]$_3$[Cu(SCN)$_4$] studied in this work.